\documentstyle[12pt]{article}
\begin{document}
\title{\bf Unbiased estimation of multi-fractal dimensions of finite data sets}
\author{A.J.~Roberts \and A.~Cronin\thanks{Both at the Department of
Mathematics \& Computing, University of Southern Queensland, Toowoomba,
Queensland 4350, Australia.  E-mail: {\tt aroberts@usq.edu.au} and {\tt
cronin@usq.edu.au}}}
\date{February 1, 1996}
\maketitle

\begin{abstract}
We present a novel method for determining multi-fractal properties from
experimental data.  It is based on maximising the likelihood that the given
finite data set comes from a particular set of parameters in a
multi-parameter family of well known multi-fractals.  By comparing
characteristic correlations obtained from the original data with those that
occur in artificially generated multi-fractals with the {\em same} number
of data points, we expect that predicted multi-fractal properties are
unbiased by the finiteness of the experimental data.
\end{abstract}


\section{Introduction}

The characterisation of spatial distributions in terms of fractal concepts
\cite{Mandelbrot79,Feder88} is becoming increasingly important.  In
particular, many distributions in nature are found to have the
characteristics of a multi-fractal\cite{Halsey86,Paladin87}: among many
examples are galaxy clustering \cite{Borgani93,Martinez91}, strange
attractors \cite{Procaccia88a}, fluid turbulence \cite{Sreenivasan91}, and
plant distributions \cite{Emmerson95}.

In application, methods for estimating fractal dimensions are typically
unreliable through many problems.  Two sources of error lie in
largely unknown biases introduced by the finiteness of data sets, addressed
by Grassberger \cite{Grassberger88b}, and in the finite range of
length-scales inherent in gathered data.  In many situations where
thousands or tens of thousands of data points are known the biases are
relatively minor; however, in some interesting problems, for example in the
spatial clustering of underwater plants \cite{Emmerson95}, only the order
of 100 data points are known and confidence in the fractal results may be
misplaced.

We propose to eliminate such biases
through a novel method of determining a multi-fractal properties of a
given dataset.  We compare characteristics of inter-point distances in the
data, \S\ref{scor}, and in artificially generated multi-fractals.  By
maximising the likelihood, \S\ref{smaxl}, that the characteristics are the
same we may model the multi-fractal nature of the data by the parameters of
the artificial multi-fractal.  By searching among, \S\ref{sopt}, artificial
multi-fractals with the same number of sample points as in the data, we
anticipate that biases due to the finite sample size will be statistically
the same in the data and in the artificial multi-fractals, \S\ref{snum};
hence predictions should be unbiased.  The method also appears to give a
reliable indication of the error in the estimates---a very desirable
feature as also noted by Judd \& Mees \cite{Judd91}.

\section{Correlation density}
\label{scor}

Borgani {\em et al} \cite{Borgani93} have surveyed the efficacy of the most
popular methods for determining fractal dimensions.  They considered:
generalised correlation integrals, box counting algorithms, density
reconstruction procedures, nearest neighbour methods, and minimal spanning
tree methods.  Their conclusion is that nearest neighbour and minimal
spanning tree methods are inferior.  Aesthetically this is pleasing because
both of these methods discard much of the available information about
inter-point distances that the other methods retain.

Box counting algorithms are very similar to the correlation integrals
except that they are non-isotropic.  Aesthetically, an isotropic method
is preferred and so here we ignore box counting methods.

Borgani {\em et al} also conclude that correlation integrals are most
reliable for generalised dimensions, $D_q$, of positive order, $q>0$, and
less reliable for negative order.  In contrast, they maintain that the
density reconstruction procedure is more reliable for negative order,
$q<0$, and less so for positive order.  The same conclusions are also
reached by Mart\'inez \cite{Martinez91}.  Interestingly, both approaches
are based on the same processed data; they just fit straight lines in
complementary manners.

Given a finite set of $N$ data points $\vec x_i$ sampled from some spatial
distribution, define the correlation functions $C_i(r)$ as the fraction
of points within a distance $r$ of the $i$th data point.  Then the
partition function
\begin{equation}
        Z(r,q)=\frac{1}{N}\sum_{i=1}^N C_i(r)^{q-1}\,,
        \label{eqz}
\end{equation}
is expected to scale as $r^\tau$ over the range of length-scales on which
the multi-fractal properties hold.  By fitting such a power law to
$Z(r,q)$, the generalised dimension $D_q$ is
then found from
\begin{equation}
        \tau=(q-1)D_q\,.
        \label{eqdq}
\end{equation}
For example, the Hausdorff dimension is estimated with $q=0$, the
correlation dimension with $q=2$, and the information dimension with
$q\to1$.

Conversely, the density reconstruction method is based on
$R_i(c)$ which is the radius of the smallest ball centred on the $i$th
point that contains a fraction $c$ of the data points.  Then the
partition function
\begin{equation}
        W(c,\tau)=\frac{1}{N}\sum_{i=1}^N R_i(c)^{-\tau}\,,
        \label{eqw}
\end{equation}
is expected to scale as $c^{1-q}$ over a respectable range of fractions
$c$.  Then by fitting a power law, $D_q$ may be again estimated
from~(\ref{eqdq}).

At the heart of both of these methods are the curves $C_i(r)$ and $R_i(c)$
which are precisely the inverse function of each other.  Imagine plotting
all of the curves, $i=1,\ldots,N$, in the $rc$-plane, and observe that:
\begin{itemize}
        \item the correlation integral $Z(r,q)$ is just an average of these
        curves over $c$ at fixed $r$;

        \item  whereas in the density reconstruction $W(c,\tau)$ is an
average of
        the curves over $r$ at fixed $c$.
\end{itemize}
One method is reputably good for positive~$q$, the other for negative~$q$.
Perhaps by using the data in the $rc$-plane directly, we can avoid such a
dichotomy between positive and negative~$q$.

\begin{figure}
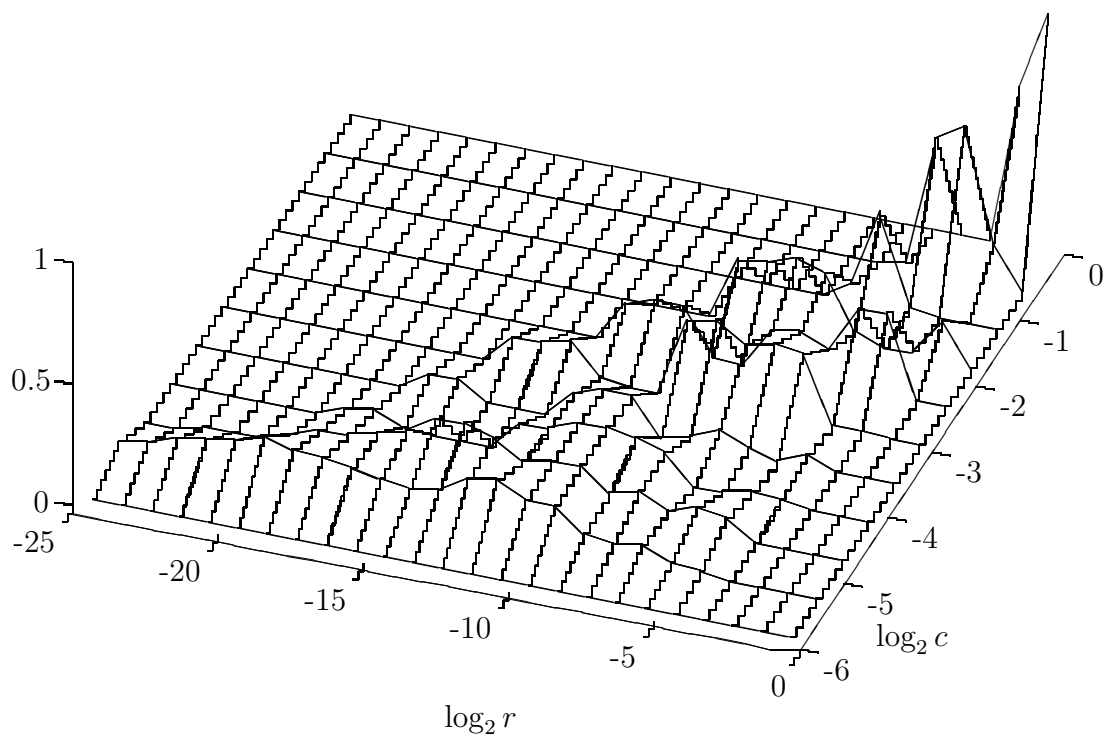

\setlength{\unitlength}{.07pt}

\caption{perspective view of the correlation density $p(r,c)$, as a
        function of $\log r$ and $\log c$, for a sample of $N=100$ points
from the
        artificially generated binary multiplicative multi-fractal shown in
        Figure~\protect\ref{fdist} with parameters $\rho=1/3$, $\phi=1/4$.}
\protect\label{fpdf}
\end{figure}

We propose to base our method on the correlation density, $p(r,c)$, defined
as being proportional to the number of $C_i(r)$ curves (or equivalently the
$R_i(c)$ curves) which pass near the point $(r,c)$.  In practise we divide
the $\log r,\log c$-plane into rectangular bins of width $\Delta r$ and
$\Delta c$, and compute $p_{jk}=p(r_j,c_k)$ as the fraction of curves with
$\log c_k-\Delta c/2<\log C_i(r_j)<\log c_k+\Delta c/2$.  See
Figure~\ref{fpdf} for example.

To emphasise that the previous two good methods are both based on this
same picture, observe that to ${\cal O}\left(\Delta^2\right)$:
\begin{equation}
        Z(r_k,q)=\sum_j p_{jk}c_j^{q-1}\,,
        \label{eqzp}
\end{equation}
whereas
\begin{equation}
        W(c_j,\tau)=\sum_k p_{jk}r_k^{-\tau}\,.
        \label{eqcp}
\end{equation}

\section{Maximum likelihood multi-fractal}
\label{smaxl}

The task now is to fit the correlation density $p(r,c)$ in a manner so that
subsequently we deduce a multi-fractal spectrum for the original data.  Our
idea is to determine parameters of a multiplicative multi-fractal which
best fits the correlation density of the original data.  Then our estimate
of the multi-fractal spectrum of the data is that of the multiplicative
multi-fractal.

The best fit is determined by:
\begin{itemize}
\item generating $N$ points on a random multiplicative multi-fractals with
a given set of parameters;

\item constructing its correlation density $P_{jk}=P(r_j,c_k)$ on the
same grid in the $rc$-plane as used in approximating $p(r,c)$ for the
original data;

\item estimating the likelihood that the two densities, $p$ and $P$, come
from the same distribution;

\item and seeking to alter the parameters of the multiplicative
multi-fractal in order to maximise this likelihood.
\end{itemize}
By probing the structure of the $N$ data points by artificially generated
fractals with precisely the same number of points we hope to eliminate, or at
least reduce, biases introduced by the finiteness of the sample size.

For definiteness, consider a sample of $N$ data points $x_i$ from a
multi-fractal on the interval $[0,1]$ in one-dimension.

\begin{figure}
\setlength{\unitlength}{.07pt}
\begin{picture}(5753,4996)
\newcommand{\w}[5]{
\put(#1,#2){\line(#3,#4){#5}}
}
\newcommand{\mh}[7]{
\multiput(#1,#2)(#3,#4){#5}{\line(#6,0){#7}}
}
\newcommand{\mv}[7]{
\multiput(#1,#2)(#3,#4){#5}{\line(0,#6){#7}}
}
\w{4687}{385}{-1}{0}{4224}
\w{4687}{4608}{-1}{0}{4224}
\w{4687}{4608}{0}{-1}{4223}
\w{463}{4608}{0}{-1}{4223}
\w{4687}{385}{1}{0}{0}
\w{463}{385}{1}{0}{0}
\w{4687}{385}{-1}{0}{4224}
\w{463}{4608}{0}{-1}{4223}
\w{463}{385}{1}{0}{0}
\w{463}{427}{0}{-1}{42}
\w{463}{4566}{0}{1}{42}
\put(423,216){0}
\w{1308}{427}{0}{-1}{42}
\w{1308}{4566}{0}{1}{42}
\put(1208,216){0.2}
\w{2153}{427}{0}{-1}{42}
\w{2153}{4566}{0}{1}{42}
\put(2053,216){0.4}
\w{2997}{427}{0}{-1}{42}
\w{2997}{4566}{0}{1}{42}
\put(2897,216){0.6}
\w{3842}{427}{0}{-1}{42}
\w{3842}{4566}{0}{1}{42}
\put(3742,216){0.8}
\w{4687}{427}{0}{-1}{42}
\w{4687}{4566}{0}{1}{42}
\put(4647,216){1}
\w{505}{385}{-1}{0}{42}
\w{4645}{385}{1}{0}{42}
\put(348,332){0}
\w{505}{807}{-1}{0}{42}
\w{4645}{807}{1}{0}{42}
\put(228,754){0.1}
\w{505}{1230}{-1}{0}{42}
\w{4645}{1230}{1}{0}{42}
\put(228,1177){0.2}
\w{505}{1652}{-1}{0}{42}
\w{4645}{1652}{1}{0}{42}
\put(228,1599){0.3}
\w{505}{2074}{-1}{0}{42}
\w{4645}{2074}{1}{0}{42}
\put(228,2021){0.4}
\w{505}{2496}{-1}{0}{42}
\w{4645}{2496}{1}{0}{42}
\put(228,2443){0.5}
\w{505}{2919}{-1}{0}{42}
\w{4645}{2919}{1}{0}{42}
\put(228,2866){0.6}
\w{505}{3341}{-1}{0}{42}
\w{4645}{3341}{1}{0}{42}
\put(228,3288){0.7}
\w{505}{3763}{-1}{0}{42}
\w{4645}{3763}{1}{0}{42}
\put(228,3710){0.8}
\w{505}{4186}{-1}{0}{42}
\w{4645}{4186}{1}{0}{42}
\put(228,4133){0.9}
\w{505}{4608}{-1}{0}{42}
\w{4645}{4608}{1}{0}{42}
\put(348,4555){1}
\w{4687}{4608}{-1}{0}{4224}
\w{4687}{385}{-1}{0}{4224}
\w{463}{4608}{0}{-1}{4223}
\w{4687}{4608}{0}{-1}{4223}
\w{463}{4608}{1}{0}{0}
\w{4687}{4608}{1}{0}{0}
\w{1331}{469}{4}{1}{181}
\mh{1512}{514}{32.5}{20}{2}{1}{32.5}
\mv{1544}{514}{32.5}{20}{2}{1}{20}
\mh{1577}{554}{4}{42}{1}{1}{4}
\mv{1581}{554}{4}{42}{1}{1}{42}
\mh{1581}{596}{5}{42}{1}{1}{5}
\mv{1586}{596}{5}{42}{1}{1}{42}
\mh{1586}{638}{16}{43}{1}{1}{16}
\mv{1602}{638}{16}{43}{1}{1}{43}
\mh{1602}{681}{1}{42}{1}{1}{1}
\mv{1603}{681}{1}{42}{1}{1}{42}
\mh{1603}{723}{1}{42}{1}{1}{1}
\mv{1604}{723}{1}{42}{1}{1}{42}
\w{1604}{765}{0}{1}{42}
\w{1604}{807}{0}{1}{43}
\mh{1604}{850}{7}{42}{1}{1}{7}
\mv{1611}{850}{7}{42}{1}{1}{42}
\w{1611}{892}{0}{1}{42}
\w{1611}{934}{0}{1}{42}
\mh{1611}{976}{1}{42}{1}{1}{1}
\mv{1612}{976}{1}{42}{1}{1}{42}
\mh{1612}{1018}{1}{43}{1}{1}{1}
\mv{1613}{1018}{1}{43}{1}{1}{43}
\mh{1613}{1061}{1}{42}{1}{1}{1}
\mv{1614}{1061}{1}{42}{1}{1}{42}
\w{1614}{1103}{6}{1}{146}
\w{1760}{1127}{1}{0}{145}
\w{1905}{1127}{6}{1}{145}
\mh{2050}{1151}{45}{18}{2}{1}{45}
\mv{2095}{1151}{45}{18}{2}{1}{18}
\w{2140}{1187}{0}{1}{43}
\w{2140}{1230}{4}{1}{166}
\mh{2306}{1271}{14}{43}{1}{1}{14}
\mv{2320}{1271}{14}{43}{1}{1}{43}
\mh{2320}{1314}{34}{21}{2}{1}{34}
\mv{2354}{1314}{34}{21}{2}{1}{21}
\mh{2388}{1356}{33}{43}{1}{1}{33}
\mv{2421}{1356}{33}{43}{1}{1}{43}
\w{2421}{1399}{1}{0}{155}
\w{2576}{1399}{6}{1}{153}
\w{2729}{1424}{1}{0}{153}
\w{2882}{1424}{6}{1}{153}
\mh{3035}{1449}{2}{34}{1}{1}{2}
\mv{3037}{1449}{2}{34}{1}{1}{34}
\mh{3037}{1483}{7}{42}{1}{1}{7}
\mv{3044}{1483}{7}{42}{1}{1}{42}
\mh{3044}{1525}{28.5}{21}{2}{1}{28.5}
\mv{3072}{1525}{28.5}{21}{2}{1}{21}
\w{3101}{1567}{0}{1}{43}
\mh{3101}{1610}{2}{42}{1}{1}{2}
\mv{3103}{1610}{2}{42}{1}{1}{42}
\mh{3103}{1652}{1}{42}{1}{1}{1}
\mv{3104}{1652}{1}{42}{1}{1}{42}
\mh{3104}{1694}{6}{42}{1}{1}{6}
\mv{3110}{1694}{6}{42}{1}{1}{42}
\w{3110}{1736}{0}{1}{43}
\mh{3110}{1779}{15}{42}{1}{1}{15}
\mv{3125}{1779}{15}{42}{1}{1}{42}
\mh{3125}{1821}{1}{42}{1}{1}{1}
\mv{3126}{1821}{1}{42}{1}{1}{42}
\w{3126}{1863}{0}{1}{42}
\w{3126}{1905}{0}{1}{43}
\w{3126}{1948}{0}{1}{42}
\mh{3126}{1990}{1}{42}{1}{1}{1}
\mv{3127}{1990}{1}{42}{1}{1}{42}
\w{3127}{2032}{0}{1}{42}
\mh{3127}{2074}{7}{42}{1}{1}{7}
\mv{3134}{2074}{7}{42}{1}{1}{42}
\w{3134}{2116}{0}{1}{43}
\mh{3134}{2159}{1}{42}{1}{1}{1}
\mv{3135}{2159}{1}{42}{1}{1}{42}
\mh{3135}{2201}{2}{42}{1}{1}{2}
\mv{3137}{2201}{2}{42}{1}{1}{42}
\w{3137}{2243}{4}{1}{145}
\w{3282}{2279}{0}{1}{49}
\mh{3282}{2328}{12}{42}{1}{1}{12}
\mv{3294}{2328}{12}{42}{1}{1}{42}
\mh{3294}{2370}{13}{42}{1}{1}{13}
\mv{3307}{2370}{13}{42}{1}{1}{42}
\w{3307}{2412}{0}{1}{42}
\mh{3307}{2454}{2}{42}{1}{1}{2}
\mv{3309}{2454}{2}{42}{1}{1}{42}
\mh{3309}{2496}{7}{43}{1}{1}{7}
\mv{3316}{2496}{7}{43}{1}{1}{43}
\w{3316}{2539}{0}{1}{42}
\mh{3316}{2581}{28}{21}{2}{1}{28}
\mv{3344}{2581}{28}{21}{2}{1}{21}
\mh{3372}{2623}{1}{42}{1}{1}{1}
\mv{3373}{2623}{1}{42}{1}{1}{42}
\w{3373}{2665}{0}{1}{43}
\w{3373}{2708}{0}{1}{42}
\mh{3373}{2750}{9}{42}{1}{1}{9}
\mv{3382}{2750}{9}{42}{1}{1}{42}
\mh{3382}{2792}{15}{42}{1}{1}{15}
\mv{3397}{2792}{15}{42}{1}{1}{42}
\w{3397}{2834}{0}{1}{43}
\w{3397}{2877}{0}{1}{42}
\w{3397}{2919}{0}{1}{42}
\w{3397}{2961}{0}{1}{42}
\w{3397}{3003}{0}{1}{42}
\w{3397}{3045}{0}{1}{43}
\w{3397}{3088}{0}{1}{42}
\w{3397}{3130}{0}{1}{42}
\w{3397}{3172}{0}{1}{42}
\w{3397}{3214}{0}{1}{43}
\mh{3397}{3257}{1}{42}{1}{1}{1}
\mv{3398}{3257}{1}{42}{1}{1}{42}
\w{3398}{3299}{0}{1}{42}
\w{3398}{3341}{0}{1}{42}
\w{3398}{3383}{0}{1}{43}
\mh{3398}{3426}{1}{42}{1}{1}{1}
\mv{3399}{3426}{1}{42}{1}{1}{42}
\w{3399}{3468}{0}{1}{42}
\w{3399}{3510}{0}{1}{42}
\w{3399}{3552}{0}{1}{42}
\w{3399}{3594}{0}{1}{43}
\mh{3399}{3637}{7}{42}{1}{1}{7}
\mv{3406}{3637}{7}{42}{1}{1}{42}
\w{3406}{3679}{0}{1}{42}
\w{3406}{3721}{0}{1}{42}
\w{3406}{3763}{0}{1}{43}
\mh{3406}{3806}{1}{42}{1}{1}{1}
\mv{3407}{3806}{1}{42}{1}{1}{42}
\mh{3407}{3848}{2}{42}{1}{1}{2}
\mv{3409}{3848}{2}{42}{1}{1}{42}
\w{3409}{3890}{6}{1}{148}
\w{3557}{3914}{1}{0}{148}
\w{3705}{3914}{6}{1}{148}
\mh{3853}{3938}{34.5}{18.5}{2}{1}{34.5}
\mv{3888}{3938}{34.5}{18.5}{2}{1}{18.5}
\mh{3922}{3975}{13}{42}{1}{1}{13}
\mv{3935}{3975}{13}{42}{1}{1}{42}
\mh{3935}{4017}{1}{42}{1}{1}{1}
\mv{3936}{4017}{1}{42}{1}{1}{42}
\w{3936}{4059}{6}{1}{246}
\mh{4182}{4100}{1}{43}{1}{1}{1}
\mv{4183}{4100}{1}{43}{1}{1}{43}
\mh{4183}{4143}{1}{43}{1}{1}{1}
\mv{4184}{4143}{1}{43}{1}{1}{43}
\mh{4184}{4186}{23}{42}{1}{1}{23}
\mv{4207}{4186}{23}{42}{1}{1}{42}
\w{4207}{4228}{0}{1}{42}
\w{4207}{4270}{0}{1}{42}
\w{4207}{4312}{0}{1}{43}
\mh{4207}{4355}{1}{42}{1}{1}{1}
\mv{4208}{4355}{1}{42}{1}{1}{42}
\w{4208}{4397}{0}{1}{42}
\w{4208}{4439}{0}{1}{42}
\w{4208}{4481}{0}{1}{43}
\mh{4208}{4524}{8}{42}{1}{1}{8}
\mv{4216}{4524}{8}{42}{1}{1}{42}
\mh{4216}{4566}{2}{42}{1}{1}{2}
\mv{4218}{4566}{2}{42}{1}{1}{42}
\w{4218}{4608}{1}{0}{469}
\w{463}{385}{1}{0}{147}
\w{610}{385}{6}{1}{144}
\w{754}{409}{1}{0}{144}
\w{898}{409}{1}{0}{144}
\w{1042}{409}{1}{0}{144}
\w{1186}{409}{6}{1}{144}
\mh{1330}{433}{0.1}{3.6}{10}{1}{0.1}
\mv{1330}{433}{0.1}{3.6}{10}{1}{3.6}
\put(2792,80){x}
\put(8,2316){M}
\w{4207}{2496}{0}{1}{10}
\w{3282}{2496}{0}{1}{10}
\w{3126}{2496}{0}{1}{10}
\w{3398}{2496}{0}{1}{10}
\w{2306}{2496}{0}{1}{10}
\w{2388}{2496}{0}{1}{10}
\w{3406}{2496}{0}{1}{10}
\w{3397}{2496}{0}{1}{10}
\w{3373}{2496}{0}{1}{10}
\w{4207}{2496}{0}{1}{10}
\w{3935}{2496}{0}{1}{10}
\w{3399}{2496}{0}{1}{10}
\w{4207}{2496}{0}{1}{10}
\w{4216}{2496}{0}{1}{10}
\w{4184}{2496}{0}{1}{10}
\w{3307}{2496}{0}{1}{10}
\w{2140}{2496}{0}{1}{10}
\w{3137}{2496}{0}{1}{10}
\w{3316}{2496}{0}{1}{10}
\w{3397}{2496}{0}{1}{10}
\w{2050}{2496}{0}{1}{10}
\w{3399}{2496}{0}{1}{10}
\w{3397}{2496}{0}{1}{10}
\w{1603}{2496}{0}{1}{10}
\w{2140}{2496}{0}{1}{10}
\w{3294}{2496}{0}{1}{10}
\w{3397}{2496}{0}{1}{10}
\w{3126}{2496}{0}{1}{10}
\w{3135}{2496}{0}{1}{10}
\w{3307}{2496}{0}{1}{10}
\w{3372}{2496}{0}{1}{10}
\w{1614}{2496}{0}{1}{10}
\w{3134}{2496}{0}{1}{10}
\w{1581}{2496}{0}{1}{10}
\w{1331}{2496}{0}{1}{10}
\w{3397}{2496}{0}{1}{10}
\w{1611}{2496}{0}{1}{10}
\w{3373}{2496}{0}{1}{10}
\w{1613}{2496}{0}{1}{10}
\w{3044}{2496}{0}{1}{10}
\w{3397}{2496}{0}{1}{10}
\w{4208}{2496}{0}{1}{10}
\w{3406}{2496}{0}{1}{10}
\w{1577}{2496}{0}{1}{10}
\w{2421}{2496}{0}{1}{10}
\w{3110}{2496}{0}{1}{10}
\w{4208}{2496}{0}{1}{10}
\w{3037}{2496}{0}{1}{10}
\w{4208}{2496}{0}{1}{10}
\w{3406}{2496}{0}{1}{10}
\w{3397}{2496}{0}{1}{10}
\w{3035}{2496}{0}{1}{10}
\w{1604}{2496}{0}{1}{10}
\w{3316}{2496}{0}{1}{10}
\w{3382}{2496}{0}{1}{10}
\w{3406}{2496}{0}{1}{10}
\w{3399}{2496}{0}{1}{10}
\w{3127}{2496}{0}{1}{10}
\w{3399}{2496}{0}{1}{10}
\w{3407}{2496}{0}{1}{10}
\w{4218}{2496}{0}{1}{10}
\w{3101}{2496}{0}{1}{10}
\w{3309}{2496}{0}{1}{10}
\w{3397}{2496}{0}{1}{10}
\w{1611}{2496}{0}{1}{10}
\w{3398}{2496}{0}{1}{10}
\w{1586}{2496}{0}{1}{10}
\w{3127}{2496}{0}{1}{10}
\w{4182}{2496}{0}{1}{10}
\w{3398}{2496}{0}{1}{10}
\w{2320}{2496}{0}{1}{10}
\w{3282}{2496}{0}{1}{10}
\w{3126}{2496}{0}{1}{10}
\w{1512}{2496}{0}{1}{10}
\w{3853}{2496}{0}{1}{10}
\w{3134}{2496}{0}{1}{10}
\w{1612}{2496}{0}{1}{10}
\w{1602}{2496}{0}{1}{10}
\w{1604}{2496}{0}{1}{10}
\w{3397}{2496}{0}{1}{10}
\w{3110}{2496}{0}{1}{10}
\w{3397}{2496}{0}{1}{10}
\w{3398}{2496}{0}{1}{10}
\w{1611}{2496}{0}{1}{10}
\w{4208}{2496}{0}{1}{10}
\w{3399}{2496}{0}{1}{10}
\w{3104}{2496}{0}{1}{10}
\w{1330}{2496}{0}{1}{10}
\w{3103}{2496}{0}{1}{10}
\w{3922}{2496}{0}{1}{10}
\w{3125}{2496}{0}{1}{10}
\w{3936}{2496}{0}{1}{10}
\w{3373}{2496}{0}{1}{10}
\w{4183}{2496}{0}{1}{10}
\w{3126}{2496}{0}{1}{10}
\w{4207}{2496}{0}{1}{10}
\w{3409}{2496}{0}{1}{10}
\w{3101}{2496}{0}{1}{10}
\w{3397}{2496}{0}{1}{10}
\end{picture}
        \caption{across the middle of this graph is plotted $N=100$ points
on an
        artificially generated binary multiplicative multi-fractal with
        parameters $\rho=1/3$ and $\phi=1/4$.  The solid line is the cumulative
        mass distribution on the fractal, $M(x)=\mbox{mass in }[0,x]$, showing
        regions of high density by large jumps in $M$.}
        \protect\label{fdist}
\end{figure}
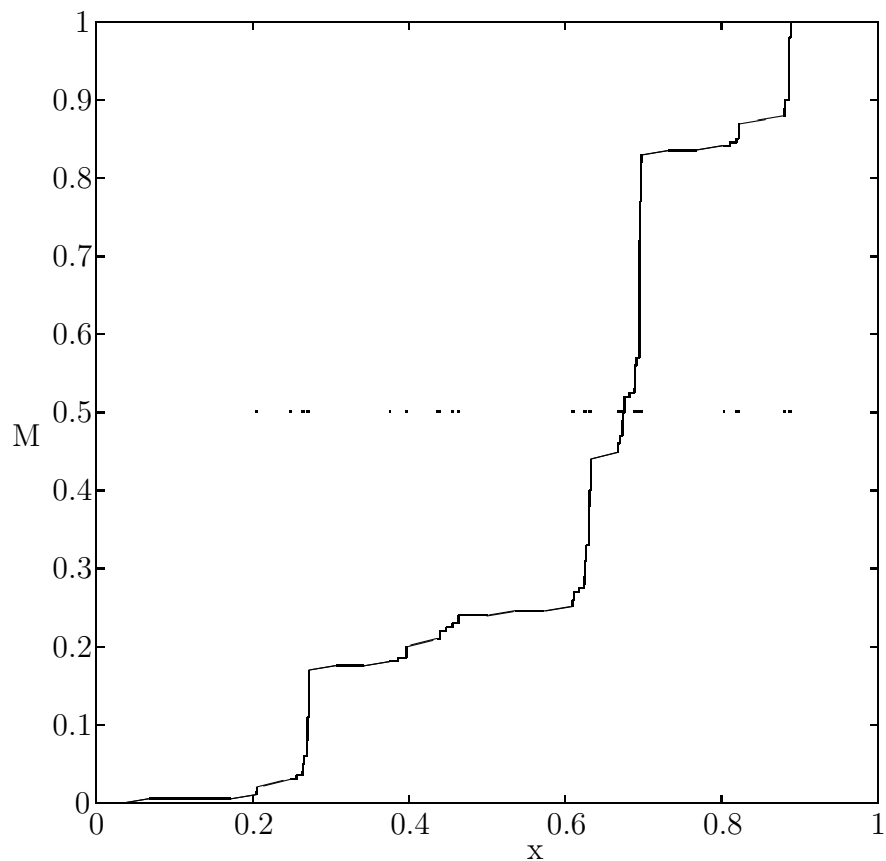

We fit from a two parameter family of binary multiplicative multi-fractals.
Given parameters $\rho\in[0,0.5]$ and $\phi\in[0,0.5]$ a binary
multiplicative multi-fractal is generated by the recursive procedure of
dividing each interval into two halves, then assigning a fraction $\phi$ of
the points in the interval to a random sub-interval of length $\rho$ in the
left
half, and the complementary fraction $\phi'=1-\phi$ to a random sub-interval of
length $\rho$ in the right half.  See, for example, the distribution shown
in Figure~\ref{fdist}.
Such a binary multiplicative multi-fractal has spectrum $f(\alpha)$
\cite[\S4]{Halsey86} given parameterically in terms of $0<\xi<1$ and
$\xi'=1-\xi$ as
\begin{equation}
                f = \frac{\xi\log{\xi}+\xi'\log{\xi'}}
                        {\log{\rho}}
                \,, \qquad
                \alpha = \frac{\xi\log{\phi}+\xi'\log{\phi'}}
                        {\log{\rho}} \,.
        \label{emfs}
\end{equation}
Other multi-fractal properties, such as generalised dimensions, then also
follow, for example, the Hausdorff dimension is just $D_0=\log
2/\log(1/\rho)$.

The likelihood that two correlation densities come from the same
distribution is determined from the $\chi^2$ statistic
\cite[\S14.3]{Press92}
\begin{equation}
        \chi^2=\sum_{p_{jk}+P_{jk}>0}\frac{(p_{jk}-P_{jk})^2}{p_{jk}+P_{jk}}\,.
        \label{echi2}
\end{equation}
Since the likelihood is monotonic in $\chi^2$ (for fixed number of degrees
of freedom), maximising likelihood is equivalent to minimising $\chi^2$.
Of course there are correlations in the structure of the correlation
densities $p(r,c)$; for example, the $C_i(r)$ curves are monotonic.
Strictly speaking these correlations destroy the precise applicability of
the likelihood and the $\chi^2$ statistic.  However, as is typical in
estimating fractal properties, such correlations are intrinsic to
investigating structures over many different length scales and are not
expected to have a serious effect.

\section{Optimum fit}
\label{sopt}

We then seek to find values of $\rho$ and $\phi$ such that the $\chi^2$
comparison between the actual data and the artificially generated data is
minimum.  Unfortunately, $\chi^2(\rho,\phi)$ has a large amount of noise due
to the random choices in the generation of the artificial binary
multiplicative multi-fractals.  Such randomness is absolutely necessary
because it characterises the wide range of possible multi-fractals with the
same multi-fractal spectrum.  (It would only be possible to eliminate the
noise if we knew a precise analytic expression for the correlation density
$P(r,c)$ for a binary multiplicative multi-fractal with parameters $\rho$
and $\phi$ {\em and} when sampled by $N$ points.) As is reasonable, the
noise in $\chi^2(\rho,\phi)$ seems to be particularly prominent for small
sample sizes $N$.

\begin{figure}
\setlength{\unitlength}{.07pt}
\begin{picture}(5813,5006)
\newcommand{\w}[5]{
\put(#1,#2){\line(#3,#4){#5}}
}
\newcommand{\mh}[7]{
\multiput(#1,#2)(#3,#4){#5}{\line(#6,0){#7}}
}
\newcommand{\mv}[7]{
\multiput(#1,#2)(#3,#4){#5}{\line(0,#6){#7}}
}
\w{4782}{395}{-1}{0}{4224}
\w{4782}{4618}{-1}{0}{4224}
\w{4782}{4618}{0}{-1}{4223}
\w{558}{4618}{0}{-1}{4223}
\w{4782}{395}{1}{0}{0}
\w{558}{395}{1}{0}{0}
\w{4782}{395}{-1}{0}{4224}
\w{558}{4618}{0}{-1}{4223}
\w{558}{395}{1}{0}{0}
\w{558}{437}{0}{-1}{42}
\w{558}{4576}{0}{1}{42}
\put(518,226){0}
\w{1403}{437}{0}{-1}{42}
\w{1403}{4576}{0}{1}{42}
\put(1303,226){0.1}
\w{2248}{437}{0}{-1}{42}
\w{2248}{4576}{0}{1}{42}
\put(2148,226){0.2}
\w{3092}{437}{0}{-1}{42}
\w{3092}{4576}{0}{1}{42}
\put(2992,226){0.3}
\w{3937}{437}{0}{-1}{42}
\w{3937}{4576}{0}{1}{42}
\put(3837,226){0.4}
\w{4782}{437}{0}{-1}{42}
\w{4782}{4576}{0}{1}{42}
\put(4682,226){0.5}
\w{600}{395}{-1}{0}{42}
\w{4740}{395}{1}{0}{42}
\put(443,342){0}
\w{600}{817}{-1}{0}{42}
\w{4740}{817}{1}{0}{42}
\put(243,764){0.05}
\w{600}{1240}{-1}{0}{42}
\w{4740}{1240}{1}{0}{42}
\put(323,1187){0.1}
\w{600}{1662}{-1}{0}{42}
\w{4740}{1662}{1}{0}{42}
\put(243,1609){0.15}
\w{600}{2084}{-1}{0}{42}
\w{4740}{2084}{1}{0}{42}
\put(323,2031){0.2}
\w{600}{2506}{-1}{0}{42}
\w{4740}{2506}{1}{0}{42}
\put(243,2453){0.25}
\w{600}{2929}{-1}{0}{42}
\w{4740}{2929}{1}{0}{42}
\put(323,2876){0.3}
\w{600}{3351}{-1}{0}{42}
\w{4740}{3351}{1}{0}{42}
\put(243,3298){0.35}
\w{600}{3773}{-1}{0}{42}
\w{4740}{3773}{1}{0}{42}
\put(323,3720){0.4}
\w{600}{4196}{-1}{0}{42}
\w{4740}{4196}{1}{0}{42}
\put(243,4143){0.45}
\w{600}{4618}{-1}{0}{42}
\w{4740}{4618}{1}{0}{42}
\put(323,4565){0.5}
\w{4782}{4618}{-1}{0}{4224}
\w{4782}{395}{-1}{0}{4224}
\w{558}{4618}{0}{-1}{4223}
\w{4782}{4618}{0}{-1}{4223}
\w{558}{4618}{1}{0}{0}
\w{4782}{4618}{1}{0}{0}
\w{3601}{2492}{0}{1}{10}
\w{3473}{2187}{0}{1}{10}
\w{3769}{2105}{0}{1}{10}
\w{3422}{2515}{0}{1}{10}
\w{3323}{2467}{0}{1}{10}
\w{3634}{2205}{0}{1}{10}
\w{3332}{2485}{0}{1}{10}
\w{3344}{2412}{0}{1}{10}
\w{3518}{2396}{0}{1}{10}
\w{3326}{2265}{0}{1}{10}
\w{3737}{2373}{0}{1}{10}
\w{3229}{2518}{0}{1}{10}
\w{3091}{2616}{0}{1}{10}
\w{3737}{2420}{0}{1}{10}
\w{3738}{2030}{0}{1}{10}
\w{2826}{2889}{0}{1}{10}
\w{3859}{2220}{0}{1}{10}
\w{3836}{2069}{0}{1}{10}
\w{3430}{2391}{0}{1}{10}
\w{3852}{2255}{0}{1}{10}
\w{3477}{2397}{0}{1}{10}
\w{3062}{2621}{0}{1}{10}
\w{3318}{2324}{0}{1}{10}
\w{3668}{2087}{0}{1}{10}
\w{3315}{2621}{0}{1}{10}
\w{3219}{2623}{0}{1}{10}
\w{3361}{2454}{0}{1}{10}
\w{3813}{2066}{0}{1}{10}
\w{3739}{2094}{0}{1}{10}
\w{3578}{2192}{0}{1}{10}
\w{3712}{2111}{0}{1}{10}
\w{3671}{2310}{0}{1}{10}
\w{3449}{2496}{0}{1}{10}
\w{3195}{2649}{0}{1}{10}
\w{3663}{2272}{0}{1}{10}
\w{3361}{2649}{0}{1}{10}
\w{3162}{2436}{0}{1}{10}
\w{3346}{2186}{0}{1}{10}
\w{3268}{2389}{0}{1}{10}
\w{3406}{2527}{0}{1}{10}
\w{3209}{2710}{0}{1}{10}
\w{3656}{2353}{0}{1}{10}
\w{3483}{2183}{0}{1}{10}
\w{4014}{1987}{0}{1}{10}
\w{3883}{2144}{0}{1}{10}
\w{3888}{2000}{0}{1}{10}
\w{3595}{2233}{0}{1}{10}
\w{3071}{2591}{0}{1}{10}
\w{3700}{2303}{0}{1}{10}
\w{3844}{2090}{0}{1}{10}
\w{3694}{2090}{0}{1}{10}
\w{3407}{2286}{0}{1}{10}
\w{3480}{2453}{0}{1}{10}
\w{3503}{2284}{0}{1}{10}
\w{2985}{2553}{0}{1}{10}
\w{3549}{2099}{0}{1}{10}
\w{3526}{2219}{0}{1}{10}
\w{3569}{2235}{0}{1}{10}
\w{3768}{2017}{0}{1}{10}
\w{3945}{2170}{0}{1}{10}
\w{3333}{2490}{0}{1}{10}
\w{3734}{2107}{0}{1}{10}
\w{3256}{2599}{0}{1}{10}
\w{3632}{2378}{0}{1}{10}
\w{3782}{2184}{0}{1}{10}
\w{3208}{2497}{0}{1}{10}
\w{3983}{1990}{0}{1}{10}
\w{3445}{2445}{0}{1}{10}
\w{3929}{2015}{0}{1}{10}
\w{3753}{2128}{0}{1}{10}
\w{3491}{2208}{0}{1}{10}
\w{4012}{1770}{0}{1}{10}
\w{3461}{2471}{0}{1}{10}
\w{3586}{2073}{0}{1}{10}
\w{3138}{2796}{0}{1}{10}
\w{3816}{1916}{0}{1}{10}
\w{3594}{2300}{0}{1}{10}
\w{4003}{2137}{0}{1}{10}
\w{3129}{2690}{0}{1}{10}
\w{3985}{1835}{0}{1}{10}
\w{4330}{1805}{0}{1}{10}
\w{3421}{2345}{0}{1}{10}
\w{3544}{2209}{0}{1}{10}
\w{3942}{2009}{0}{1}{10}
\w{3688}{2109}{0}{1}{10}
\w{3923}{2283}{0}{1}{10}
\w{3346}{2380}{0}{1}{10}
\w{3529}{2326}{0}{1}{10}
\w{3684}{2285}{0}{1}{10}
\w{3928}{1990}{0}{1}{10}
\w{3518}{2204}{0}{1}{10}
\w{4096}{1891}{0}{1}{10}
\w{3901}{2157}{0}{1}{10}
\w{3573}{2651}{0}{1}{10}
\w{3615}{2194}{0}{1}{10}
\w{3604}{2234}{0}{1}{10}
\w{3816}{2194}{0}{1}{10}
\w{3657}{2309}{0}{1}{10}
\w{3618}{2438}{0}{1}{10}
\w{3937}{2122}{0}{1}{10}
\put(3576,2290){\circle {72}}
\w{3410}{2506}{-1}{0}{72}
\w{3374}{2470}{0}{1}{72}
\put(2881,66){$\rho$}
\put(19,2350){$\phi$}
\end{picture}
\caption{a cloud of $m=100$ parameter values which match to a good degree
of likelihood with an artificially generated binary multiplicative
multi-fractal with parameters $\rho=1/3$ and $\phi_1=1/4$.  The $+$ denotes
the actual parameters used to generate the original data sample, whereas
the $\circ$
denotes the mean of the the cloud as our best estimate of the parameters.}
\protect\label{fcloud}
\end{figure}
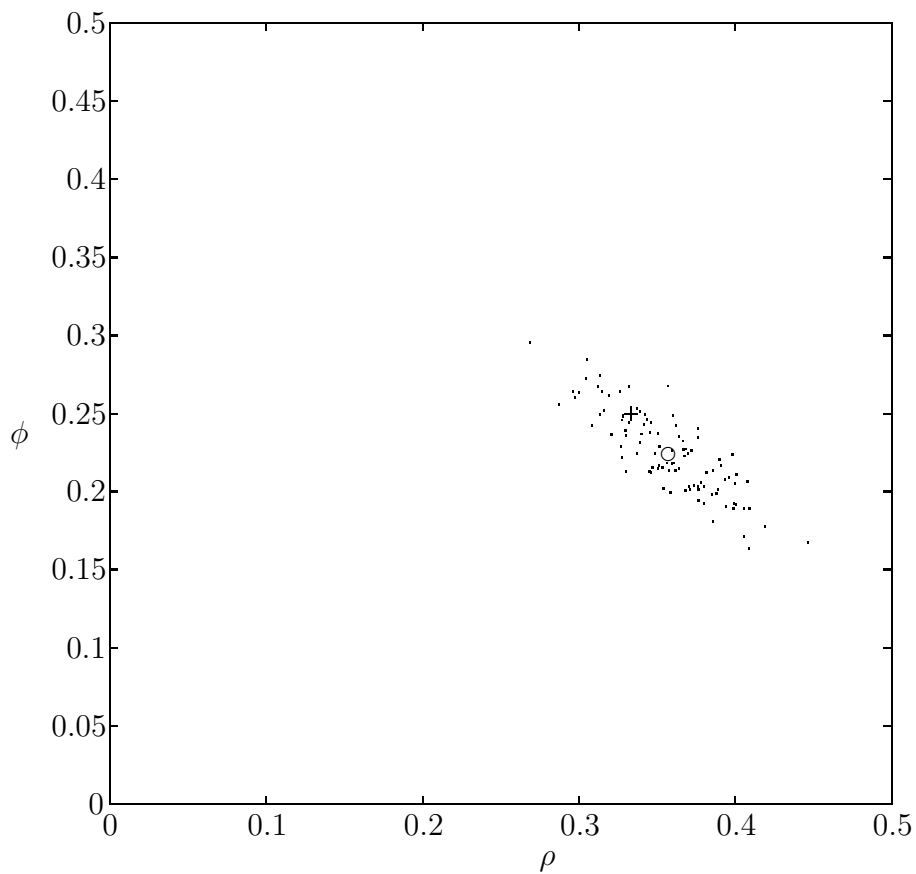

The practical task is to minimise a noisy function of $\rho$ and $\phi$.
Various approaches were tried; however, a crude procedure reminiscent of
simulated annealing seems to be effective.  We choose the $m$ parameters
$(\rho_\ell,\phi_\ell)$, typically $m=100$, of least evaluated $\chi^2$
from a sample of parameter values.  The sample of parameter values is
initially uniformly randomly distributed over parameter space,
$[0,0.5]\times[0,0.5]$, and then more are generated from the $m$ most
successful so far.  Typically 20~iterations were performed, constructing
some 2000 artificial multi-fractals and using $\chi^2$ to compare their
correlation density with the original.  Ultimately we end with a cloud of
$m$ points, as shown in Figure~\ref{fcloud}, corresponding to parameters
which have realised multi-fractals with a good fit to the correlation
density of the original data.
The centre of the cloud, the mean of the parameter points, is our best
estimate of the appropriate parameters to use in order to model the
original data.  The spread of the cloud is indicative of the sensitivity of
the parameter estimation to the noise inherent in fitting to the
correlation density given the limited amount of information
 in $N$ data points---the spread
depends upon the signal:noise ratio.

The optimisation procedure reminds us of simulated annealing in that
parameter values are retained depending upon randomness, though here the
randomness is intrinsic to the function rather than externally imposed.
The ``temperature'', which is progressively lowered in simulated annealing,
is analogous to the threshold of the $m$the best parameter value, which
decreases from one iteration to the next.

\section{Numerical experiments}
\label{snum}

One numerical trial consists of generating a multiplicative multi-fractal
with a specific value of its parameters, here we chose $\rho=1/3$ and
$\phi=1/4$ throughout, sampled with $N$ points where here we used $N=30$,
$100$ (shown previously) or $300$.  This multi-fractal forms the synthetic
finite data set.  Then the above procedure is applied and the mean of
the ultimate cloud of parameter values recorded as an estimate of the
parameters of the synthetic data.

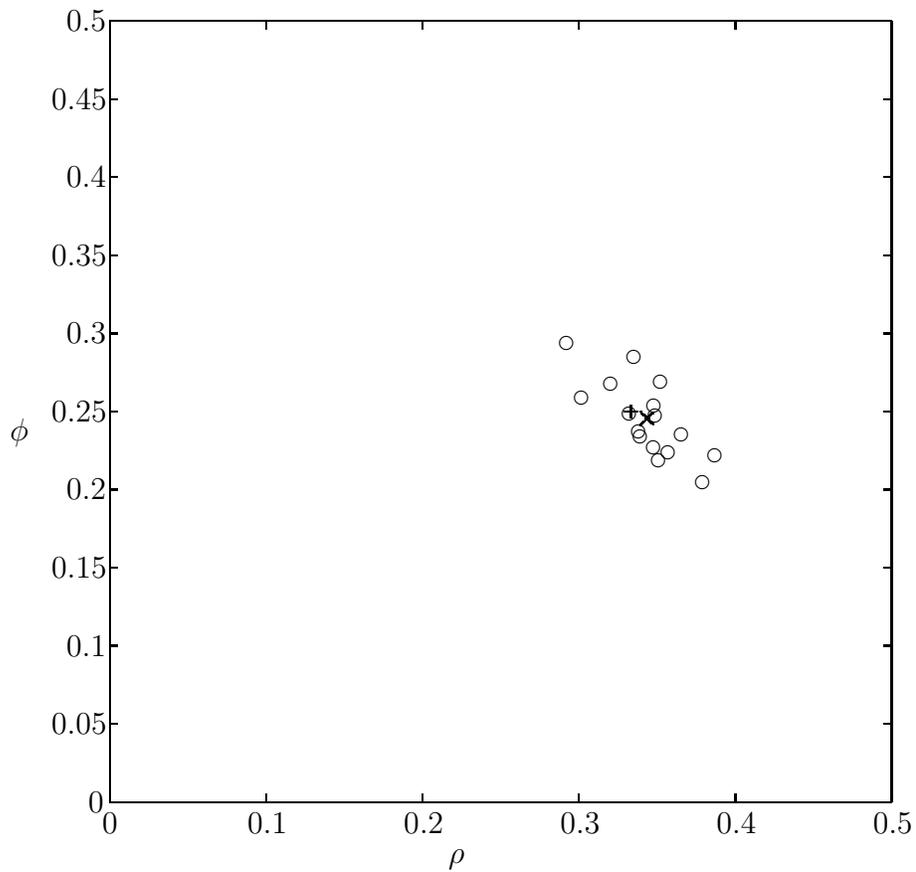
\begin{figure}
\setlength{\unitlength}{.07pt}
\begin{picture}(5813,5006)
\newcommand{\w}[5]{
\put(#1,#2){\line(#3,#4){#5}}
}
\newcommand{\mh}[7]{
\multiput(#1,#2)(#3,#4){#5}{\line(#6,0){#7}}
}
\newcommand{\mv}[7]{
\multiput(#1,#2)(#3,#4){#5}{\line(0,#6){#7}}
}
\w{4782}{395}{-1}{0}{4224}
\w{4782}{4618}{-1}{0}{4224}
\w{4782}{4618}{0}{-1}{4223}
\w{558}{4618}{0}{-1}{4223}
\w{4782}{395}{1}{0}{0}
\w{558}{395}{1}{0}{0}
\w{4782}{395}{-1}{0}{4224}
\w{558}{4618}{0}{-1}{4223}
\w{558}{395}{1}{0}{0}
\w{558}{437}{0}{-1}{42}
\w{558}{4576}{0}{1}{42}
\put(518,226){0}
\w{1403}{437}{0}{-1}{42}
\w{1403}{4576}{0}{1}{42}
\put(1303,226){0.1}
\w{2248}{437}{0}{-1}{42}
\w{2248}{4576}{0}{1}{42}
\put(2148,226){0.2}
\w{3092}{437}{0}{-1}{42}
\w{3092}{4576}{0}{1}{42}
\put(2992,226){0.3}
\w{3937}{437}{0}{-1}{42}
\w{3937}{4576}{0}{1}{42}
\put(3837,226){0.4}
\w{4782}{437}{0}{-1}{42}
\w{4782}{4576}{0}{1}{42}
\put(4682,226){0.5}
\w{600}{395}{-1}{0}{42}
\w{4740}{395}{1}{0}{42}
\put(443,342){0}
\w{600}{817}{-1}{0}{42}
\w{4740}{817}{1}{0}{42}
\put(243,764){0.05}
\w{600}{1240}{-1}{0}{42}
\w{4740}{1240}{1}{0}{42}
\put(323,1187){0.1}
\w{600}{1662}{-1}{0}{42}
\w{4740}{1662}{1}{0}{42}
\put(243,1609){0.15}
\w{600}{2084}{-1}{0}{42}
\w{4740}{2084}{1}{0}{42}
\put(323,2031){0.2}
\w{600}{2506}{-1}{0}{42}
\w{4740}{2506}{1}{0}{42}
\put(243,2453){0.25}
\w{600}{2929}{-1}{0}{42}
\w{4740}{2929}{1}{0}{42}
\put(323,2876){0.3}
\w{600}{3351}{-1}{0}{42}
\w{4740}{3351}{1}{0}{42}
\put(243,3298){0.35}
\w{600}{3773}{-1}{0}{42}
\w{4740}{3773}{1}{0}{42}
\put(323,3720){0.4}
\w{600}{4196}{-1}{0}{42}
\w{4740}{4196}{1}{0}{42}
\put(243,4143){0.45}
\w{600}{4618}{-1}{0}{42}
\w{4740}{4618}{1}{0}{42}
\put(323,4565){0.5}
\w{4782}{4618}{-1}{0}{4224}
\w{4782}{395}{-1}{0}{4224}
\w{558}{4618}{0}{-1}{4223}
\w{4782}{4618}{0}{-1}{4223}
\w{558}{4618}{1}{0}{0}
\w{4782}{4618}{1}{0}{0}
\put(3574,2287){\circle {72}}
\put(3025,2878){\circle {72}}
\put(3388,2802){\circle {72}}
\put(3760,2126){\circle {72}}
\put(3827,2270){\circle {72}}
\put(3422,2371){\circle {72}}
\put(3532,2667){\circle {72}}
\put(3363,2498){\circle {72}}
\put(3503,2485){\circle {72}}
\put(3414,2401){\circle {72}}
\put(3496,2541){\circle {72}}
\put(3262,2659){\circle {72}}
\put(3105,2581){\circle {72}}
\put(3520,2241){\circle {72}}
\put(3495,2313){\circle {72}}
\put(3644,2383){\circle {72}}
\put(2389,66){$\rho$}
\put(19,2350){$\phi$}
\w{3410}{2506}{-1}{0}{72}
\w{3374}{2470}{0}{1}{72}
\mv{3494}{2433}{-7.2}{7.2}{10}{1}{7.2}
\mh{3494}{2440}{-7.2}{7.2}{10}{-1}{7.2}
\mh{3422}{2433}{7.2}{7.2}{10}{1}{7.2}
\mv{3429}{2433}{7.2}{7.2}{10}{1}{7.2}
\end{picture}
\caption{predicted multi-fractal parameters $(\rho,\phi)$, indicated by
$\circ$'s, from the
maximum likelihood match to an ensemble of 16 different realisations, each
of $N=100$ data points, of a binary multiplicative multi-fractal with
parameters $\rho=1/3$ and $\phi=1/4$, indicated by $+$.  The mean location
of the realisations and predictions is indicated by a $\times$.}
\protect\label{n100means}
\end{figure}

To test the method, we performed 16 different trials with $N=30$ data
points, $N=100$, and $N=300$.  The estimates of the multi-fractal
parameters for $N=100$ and $N=300$ are shown in Figures~\ref{n100means}
and~\ref{n300means}, respectively.
\begin{figure}
\setlength{\unitlength}{.07pt}
\begin{picture}(6037,5006)
\newcommand{\w}[5]{
\put(#1,#2){\line(#3,#4){#5}}
}
\newcommand{\mh}[7]{
\multiput(#1,#2)(#3,#4){#5}{\line(#6,0){#7}}
}
\newcommand{\mv}[7]{
\multiput(#1,#2)(#3,#4){#5}{\line(0,#6){#7}}
}
\w{4782}{395}{-1}{0}{4224}
\w{4782}{4618}{-1}{0}{4224}
\w{4782}{4618}{0}{-1}{4223}
\w{558}{4618}{0}{-1}{4223}
\w{4782}{395}{1}{0}{0}
\w{558}{395}{1}{0}{0}
\w{4782}{395}{-1}{0}{4224}
\w{558}{4618}{0}{-1}{4223}
\w{558}{395}{1}{0}{0}
\w{558}{437}{0}{-1}{42}
\w{558}{4576}{0}{1}{42}
\put(418,226){0.25}
\w{1966}{437}{0}{-1}{42}
\w{1966}{4576}{0}{1}{42}
\put(1866,226){0.3}
\w{3374}{437}{0}{-1}{42}
\w{3374}{4576}{0}{1}{42}
\put(3234,226){0.35}
\w{4782}{437}{0}{-1}{42}
\w{4782}{4576}{0}{1}{42}
\put(4682,226){0.4}
\w{600}{395}{-1}{0}{42}
\w{4740}{395}{1}{0}{42}
\put(323,342){0.2}
\w{600}{817}{-1}{0}{42}
\w{4740}{817}{1}{0}{42}
\put(243,764){0.21}
\w{600}{1240}{-1}{0}{42}
\w{4740}{1240}{1}{0}{42}
\put(243,1187){0.22}
\w{600}{1662}{-1}{0}{42}
\w{4740}{1662}{1}{0}{42}
\put(243,1609){0.23}
\w{600}{2084}{-1}{0}{42}
\w{4740}{2084}{1}{0}{42}
\put(243,2031){0.24}
\w{600}{2507}{-1}{0}{42}
\w{4740}{2507}{1}{0}{42}
\put(243,2454){0.25}
\w{600}{2929}{-1}{0}{42}
\w{4740}{2929}{1}{0}{42}
\put(243,2876){0.26}
\w{600}{3351}{-1}{0}{42}
\w{4740}{3351}{1}{0}{42}
\put(243,3298){0.27}
\w{600}{3773}{-1}{0}{42}
\w{4740}{3773}{1}{0}{42}
\put(243,3720){0.28}
\w{600}{4196}{-1}{0}{42}
\w{4740}{4196}{1}{0}{42}
\put(243,4143){0.29}
\w{600}{4618}{-1}{0}{42}
\w{4740}{4618}{1}{0}{42}
\put(323,4565){0.3}
\w{4782}{4618}{-1}{0}{4224}
\w{4782}{395}{-1}{0}{4224}
\w{558}{4618}{0}{-1}{4223}
\w{4782}{4618}{0}{-1}{4223}
\w{558}{4618}{1}{0}{0}
\w{4782}{4618}{1}{0}{0}
\put(2388,3098){\circle {72}}
\put(3036,2042){\circle {72}}
\put(3008,2084){\circle {72}}
\put(3149,2861){\circle {72}}
\put(3143,2595){\circle {72}}
\put(3160,1691){\circle {72}}
\put(2121,3617){\circle {72}}
\put(3126,2515){\circle {72}}
\put(3137,2794){\circle {72}}
\put(3118,2397){\circle {72}}
\put(3329,2384){\circle {72}}
\put(3326,2367){\circle {72}}
\put(3357,2439){\circle {72}}
\put(3259,2629){\circle {72}}
\put(2943,2823){\circle {72}}
\put(3106,2637){\circle {72}}
\put(2389,66){$\rho$}
\put(19,2350){$\phi$}
\w{2941}{2507}{-1}{0}{72}
\w{2905}{2471}{0}{1}{72}
\mv{3080}{2525}{-7.2}{7.2}{10}{1}{7.2}
\mh{3080}{2532}{-7.2}{7.2}{10}{-1}{7.2}
\mh{3008}{2525}{7.2}{7.2}{10}{1}{7.2}
\mv{3015}{2525}{7.2}{7.2}{10}{1}{7.2}
\w{3367}{2732}{0}{1}{10}
\w{2706}{2964}{0}{1}{10}
\w{3175}{2877}{0}{1}{10}
\w{2903}{1975}{0}{1}{10}
\w{2442}{2236}{0}{1}{10}
\w{2925}{2793}{0}{1}{10}
\w{3378}{2322}{0}{1}{10}
\w{3071}{2912}{0}{1}{10}
\w{3127}{2434}{0}{1}{10}
\w{3119}{3143}{0}{1}{10}
\w{2820}{3486}{0}{1}{10}
\w{2918}{2584}{0}{1}{10}
\w{3269}{1651}{0}{1}{10}
\w{2974}{2565}{0}{1}{10}
\w{3139}{2333}{0}{1}{10}
\w{3181}{2057}{0}{1}{10}
\w{2800}{2203}{0}{1}{10}
\w{3005}{2340}{0}{1}{10}
\w{3631}{2075}{0}{1}{10}
\w{2472}{3117}{0}{1}{10}
\w{3268}{2714}{0}{1}{10}
\w{3517}{2412}{0}{1}{10}
\w{2755}{3289}{0}{1}{10}
\w{3112}{2280}{0}{1}{10}
\w{3374}{2373}{0}{1}{10}
\w{3183}{3132}{0}{1}{10}
\w{4287}{1427}{0}{1}{10}
\w{2746}{3175}{0}{1}{10}
\w{2360}{2460}{0}{1}{10}
\w{3048}{1889}{0}{1}{10}
\w{2079}{2902}{0}{1}{10}
\w{2854}{3141}{0}{1}{10}
\w{2325}{2747}{0}{1}{10}
\w{3087}{2224}{0}{1}{10}
\w{3307}{2287}{0}{1}{10}
\w{3443}{2663}{0}{1}{10}
\w{3787}{1622}{0}{1}{10}
\w{2672}{2750}{0}{1}{10}
\w{2683}{2350}{0}{1}{10}
\w{3122}{2835}{0}{1}{10}
\w{1992}{3343}{0}{1}{10}
\w{2622}{3263}{0}{1}{10}
\w{3159}{2039}{0}{1}{10}
\w{3213}{2492}{0}{1}{10}
\w{3299}{2685}{0}{1}{10}
\w{2665}{3352}{0}{1}{10}
\w{2542}{3434}{0}{1}{10}
\w{3101}{2721}{0}{1}{10}
\w{3259}{2517}{0}{1}{10}
\w{2244}{4396}{0}{1}{10}
\w{3765}{2142}{0}{1}{10}
\w{2975}{2514}{0}{1}{10}
\w{2941}{2707}{0}{1}{10}
\w{3076}{2147}{0}{1}{10}
\w{3633}{2642}{0}{1}{10}
\w{2453}{2775}{0}{1}{10}
\w{3565}{2114}{0}{1}{10}
\w{3182}{2626}{0}{1}{10}
\w{2537}{3394}{0}{1}{10}
\w{2441}{2960}{0}{1}{10}
\w{2713}{2263}{0}{1}{10}
\w{3062}{3278}{0}{1}{10}
\w{3026}{1378}{0}{1}{10}
\w{3062}{2507}{0}{1}{10}
\w{3251}{2658}{0}{1}{10}
\w{2922}{3122}{0}{1}{10}
\w{3405}{2392}{0}{1}{10}
\w{3714}{1929}{0}{1}{10}
\w{3810}{2523}{0}{1}{10}
\w{2766}{2589}{0}{1}{10}
\w{2942}{2914}{0}{1}{10}
\w{2427}{3132}{0}{1}{10}
\w{3768}{1177}{0}{1}{10}
\w{2197}{2721}{0}{1}{10}
\w{2718}{3102}{0}{1}{10}
\w{2074}{3143}{0}{1}{10}
\w{2802}{2527}{0}{1}{10}
\w{4008}{1191}{0}{1}{10}
\w{1752}{3167}{0}{1}{10}
\w{3709}{2887}{0}{1}{10}
\w{3745}{2135}{0}{1}{10}
\w{2391}{3303}{0}{1}{10}
\w{3913}{1377}{0}{1}{10}
\w{4527}{1303}{0}{1}{10}
\w{4203}{1991}{0}{1}{10}
\w{3231}{2756}{0}{1}{10}
\w{3123}{2323}{0}{1}{10}
\w{3249}{3423}{0}{1}{10}
\w{4371}{922}{0}{1}{10}
\w{3398}{1488}{0}{1}{10}
\w{3163}{3636}{0}{1}{10}
\w{4459}{414}{0}{1}{10}
\w{3104}{3544}{0}{1}{10}
\w{2845}{2775}{0}{1}{10}
\w{4730}{634}{0}{1}{10}
\w{2780}{3298}{0}{1}{10}
\w{3587}{2530}{0}{1}{10}
\w{3164}{2353}{0}{1}{10}
\w{3409}{2483}{0}{1}{10}
\end{picture}
\caption{predicted
multi-fractal parameters $(\rho,\phi)$, indicated by $\circ$'s, from the
maximum likelihood match to an emsemble of 16 different realisations, each
of $N=300$ data points, of a binary multiplicative multi-fractal with
parameters $\rho=1/3$ and $\phi=1/4$, indicated by $+$.  The mean location
of the realisations and predictions is indicated by a $\times$.  A cloud of
best likelihood comparisons for one of the realisations is shown by the
dots.  Note the change in scale for this Figure.}
\protect\label{n300means}
\end{figure}
Observe that, as would be expected, the parameter predictions improve for
the larger value of $N$.  Also observe that the numerical predictions of
$(\rho,\phi)$ are reasonably centred upon the true values of $(1/3,1/4)$.
The method does indeed seem to lack any bias due to the finite size of the
sample.

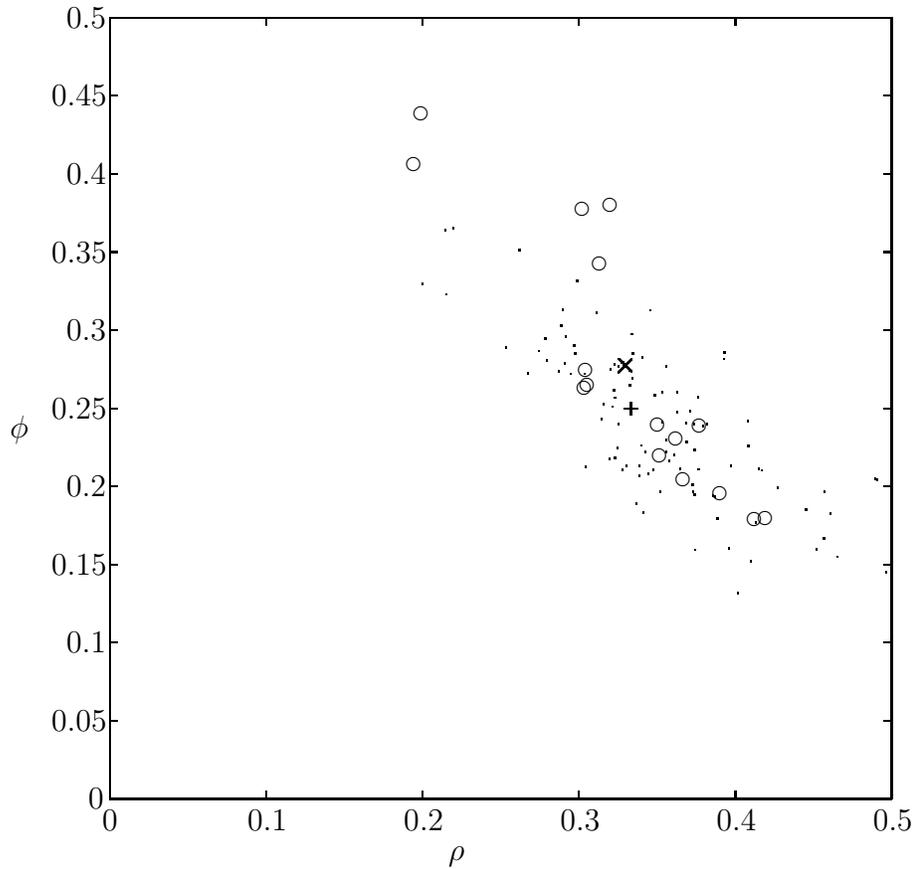
\begin{figure}
\setlength{\unitlength}{.07pt}
\begin{picture}(5813,5006)
\newcommand{\w}[5]{
\put(#1,#2){\line(#3,#4){#5}}
}
\newcommand{\mh}[7]{
\multiput(#1,#2)(#3,#4){#5}{\line(#6,0){#7}}
}
\newcommand{\mv}[7]{
\multiput(#1,#2)(#3,#4){#5}{\line(0,#6){#7}}
}
\w{4782}{395}{-1}{0}{4224}
\w{4782}{4618}{-1}{0}{4224}
\w{4782}{4618}{0}{-1}{4223}
\w{558}{4618}{0}{-1}{4223}
\w{4782}{395}{1}{0}{0}
\w{558}{395}{1}{0}{0}
\w{4782}{395}{-1}{0}{4224}
\w{558}{4618}{0}{-1}{4223}
\w{558}{395}{1}{0}{0}
\w{558}{437}{0}{-1}{42}
\w{558}{4576}{0}{1}{42}
\put(518,226){0}
\w{1403}{437}{0}{-1}{42}
\w{1403}{4576}{0}{1}{42}
\put(1303,226){0.1}
\w{2248}{437}{0}{-1}{42}
\w{2248}{4576}{0}{1}{42}
\put(2148,226){0.2}
\w{3092}{437}{0}{-1}{42}
\w{3092}{4576}{0}{1}{42}
\put(2992,226){0.3}
\w{3937}{437}{0}{-1}{42}
\w{3937}{4576}{0}{1}{42}
\put(3837,226){0.4}
\w{4782}{437}{0}{-1}{42}
\w{4782}{4576}{0}{1}{42}
\put(4682,226){0.5}
\w{600}{395}{-1}{0}{42}
\w{4740}{395}{1}{0}{42}
\put(443,342){0}
\w{600}{817}{-1}{0}{42}
\w{4740}{817}{1}{0}{42}
\put(243,764){0.05}
\w{600}{1240}{-1}{0}{42}
\w{4740}{1240}{1}{0}{42}
\put(323,1187){0.1}
\w{600}{1662}{-1}{0}{42}
\w{4740}{1662}{1}{0}{42}
\put(243,1609){0.15}
\w{600}{2084}{-1}{0}{42}
\w{4740}{2084}{1}{0}{42}
\put(323,2031){0.2}
\w{600}{2506}{-1}{0}{42}
\w{4740}{2506}{1}{0}{42}
\put(243,2453){0.25}
\w{600}{2929}{-1}{0}{42}
\w{4740}{2929}{1}{0}{42}
\put(323,2876){0.3}
\w{600}{3351}{-1}{0}{42}
\w{4740}{3351}{1}{0}{42}
\put(243,3298){0.35}
\w{600}{3773}{-1}{0}{42}
\w{4740}{3773}{1}{0}{42}
\put(323,3720){0.4}
\w{600}{4196}{-1}{0}{42}
\w{4740}{4196}{1}{0}{42}
\put(243,4143){0.45}
\w{600}{4618}{-1}{0}{42}
\w{4740}{4618}{1}{0}{42}
\put(323,4565){0.5}
\w{4782}{4618}{-1}{0}{4224}
\w{4782}{395}{-1}{0}{4224}
\w{558}{4618}{0}{-1}{4223}
\w{4782}{4618}{0}{-1}{4223}
\w{558}{4618}{1}{0}{0}
\w{4782}{4618}{1}{0}{0}
\put(3515,2422){\circle {72}}
\put(3135,2633){\circle {72}}
\put(3202,3292){\circle {72}}
\put(3109,3588){\circle {72}}
\put(3126,2718){\circle {72}}
\put(4098,1915){\circle {72}}
\put(3853,2050){\circle {72}}
\put(3743,2414){\circle {72}}
\put(2238,4104){\circle {72}}
\put(3615,2344){\circle {72}}
\put(3527,2251){\circle {72}}
\put(3119,2617){\circle {72}}
\put(3259,3607){\circle {72}}
\put(4039,1907){\circle {72}}
\put(2199,3825){\circle {72}}
\put(3655,2125){\circle {72}}
\put(2389,66){$\rho$}
\put(19,2350){$\phi$}
\w{3410}{2506}{-1}{0}{72}
\w{3374}{2470}{0}{1}{72}
\mv{3376}{2702}{-7.2}{7.2}{10}{1}{7.2}
\mh{3376}{2709}{-7.2}{7.2}{10}{-1}{7.2}
\mh{3304}{2702}{7.2}{7.2}{10}{1}{7.2}
\mv{3311}{2702}{7.2}{7.2}{10}{1}{7.2}
\w{3382}{2664}{0}{1}{10}
\w{3434}{2778}{0}{1}{10}
\w{4081}{2167}{0}{1}{10}
\w{3073}{2798}{0}{1}{10}
\w{3674}{2321}{0}{1}{10}
\w{3608}{2251}{0}{1}{10}
\w{3305}{2417}{0}{1}{10}
\w{4048}{1884}{0}{1}{10}
\w{3582}{2217}{0}{1}{10}
\w{3431}{2302}{0}{1}{10}
\w{3641}{2176}{0}{1}{10}
\w{3441}{1940}{0}{1}{10}
\w{3348}{2193}{0}{1}{10}
\w{3187}{3019}{0}{1}{10}
\w{3082}{3189}{0}{1}{10}
\w{3623}{2484}{0}{1}{10}
\w{3275}{2511}{0}{1}{10}
\w{3418}{2138}{0}{1}{10}
\w{3288}{2560}{0}{1}{10}
\w{3123}{2688}{0}{1}{10}
\w{3565}{2267}{0}{1}{10}
\w{3264}{2711}{0}{1}{10}
\w{3328}{2171}{0}{1}{10}
\w{3902}{1746}{0}{1}{10}
\w{2983}{2703}{0}{1}{10}
\w{3067}{2842}{0}{1}{10}
\w{2997}{2949}{0}{1}{10}
\w{3018}{2747}{0}{1}{10}
\w{4166}{2073}{0}{1}{10}
\w{2246}{3173}{0}{1}{10}
\w{3333}{2750}{0}{1}{10}
\w{3720}{1737}{0}{1}{10}
\w{3913}{2189}{0}{1}{10}
\w{3879}{2801}{0}{1}{10}
\w{3021}{2891}{0}{1}{10}
\w{2700}{2829}{0}{1}{10}
\w{3383}{2798}{0}{1}{10}
\w{3567}{2728}{0}{1}{10}
\w{3453}{2267}{0}{1}{10}
\w{4453}{1932}{0}{1}{10}
\w{3764}{2405}{0}{1}{10}
\w{3050}{2689}{0}{1}{10}
\w{2918}{2763}{0}{1}{10}
\w{3288}{2234}{0}{1}{10}
\w{2413}{3476}{0}{1}{10}
\w{3302}{2287}{0}{1}{10}
\w{3709}{2052}{0}{1}{10}
\w{4022}{1677}{0}{1}{10}
\w{3715}{2417}{0}{1}{10}
\w{3368}{2624}{0}{1}{10}
\w{2375}{3118}{0}{1}{10}
\w{2370}{3466}{0}{1}{10}
\w{3953}{1501}{0}{1}{10}
\w{3544}{2586}{0}{1}{10}
\w{3735}{2565}{0}{1}{10}
\w{3563}{2330}{0}{1}{10}
\w{3495}{2171}{0}{1}{10}
\w{3420}{2192}{0}{1}{10}
\w{4375}{1740}{0}{1}{10}
\w{3533}{2052}{0}{1}{10}
\w{3226}{2524}{0}{1}{10}
\w{4319}{1954}{0}{1}{10}
\w{2817}{2693}{0}{1}{10}
\w{4008}{2300}{0}{1}{10}
\w{3256}{2227}{0}{1}{10}
\w{4694}{2118}{0}{1}{10}
\w{3623}{2586}{0}{1}{10}
\w{2911}{2877}{0}{1}{10}
\w{3285}{2739}{0}{1}{10}
\w{4422}{2051}{0}{1}{10}
\w{3740}{2172}{0}{1}{10}
\w{3818}{2030}{0}{1}{10}
\w{4752}{1614}{0}{1}{10}
\w{3717}{2278}{0}{1}{10}
\w{2771}{3356}{0}{1}{10}
\w{3405}{1987}{0}{1}{10}
\w{3877}{2769}{0}{1}{10}
\w{3307}{2730}{0}{1}{10}
\w{3786}{2418}{0}{1}{10}
\w{3503}{2575}{0}{1}{10}
\w{3671}{2422}{0}{1}{10}
\w{3842}{1903}{0}{1}{10}
\w{3468}{2150}{0}{1}{10}
\w{3283}{2601}{0}{1}{10}
\w{3707}{2092}{0}{1}{10}
\w{3476}{3033}{0}{1}{10}
\w{3217}{2445}{0}{1}{10}
\w{3378}{2903}{0}{1}{10}
\w{3544}{2428}{0}{1}{10}
\w{4703}{2115}{0}{1}{10}
\w{3718}{2037}{0}{1}{10}
\w{3695}{2488}{0}{1}{10}
\w{4416}{1798}{0}{1}{10}
\w{3127}{2185}{0}{1}{10}
\w{4005}{2435}{0}{1}{10}
\w{2875}{2813}{0}{1}{10}
\w{3004}{3035}{0}{1}{10}
\w{4488}{1699}{0}{1}{10}
\w{4066}{2177}{0}{1}{10}
\w{3828}{2024}{0}{1}{10}
\end{picture}
\caption{predicted
multi-fractal parameters $(\rho,\phi)$, indicated by $\circ$'s, from the
maximum likelihood match to an ensemble of 16 different realisations, each
of $N=30$ data points, of a binary multiplicative multi-fractal with
parameters $\rho=1/3$ and $\phi=1/4$, indicated by $+$.  The mean location
of the realisations and predictions is indicated by a $\times$.  A cloud of
best likelihood comparisons for one of the realisations is shown by the
dots.}
\protect\label{n30means}
\end{figure}

Going to an extremely small number of data points, $N=30$, as shown in
Figure~\ref{n30means} we find that apart from two bad realisations (of
small $\rho$, high $\phi$) the predicted multi-fractal parameters are
semi-quantitative in that they indicate a reasonably limited area of the
parameter space in which the actual multi-fractal parameters are to be
found.  We consider this an impressive result for such a small $N$.

A useful observation is that the spread observed between the estimates in
such an ensemble of realisations, such as in Figure~\ref{n100means}, is
approximately the spread observed in the cloud of high likelihood
parameters, see a corresponding cloud in Figure~\ref{fcloud}: the
covariances are very similar.  The similar spread of the cloud of high
likelihood comparisons to the inherent error of the multi-fractal
parameter prediction is also demonstrated in Figures~\ref{n30means}
and~\ref{n300means} where the cloud for just one of the realisations is
also plotted.  Thus in application, where one generally
cannot undertake more than one sample, we expect that the cloud will be a
reliable estimate of the region within which we can confidentially expect
the true multi-fractal parameters to lie.  Indeed, it was almost always
the case that the true value of $(\rho,\phi)$ lay within the cloud of
high likelihood parameter values (the only exceptions being the two bad
fits with $N=30$ data points).  Thus the covariance matrix of the cloud
gives a good indication of the error in the fitted parameter values.

We expect that the reason the cloud gives a good estimate of the error is
that the size of the cloud is characteristic of the signal:noise ratio in
$\chi^2(\rho,\phi)$.  However, the noise comes from randomly sampling a
multi-fractal, namely the binary multiplicative multi-fractal, and hence
is similar to the bias introduced by the {\em same} sized sample of the
natural process that generated the original data.  Thus the influence of
this bias is characteristic of the influence of the noise we see in the
analysis, and so the latter may be used to predict errors.

\begin{figure}
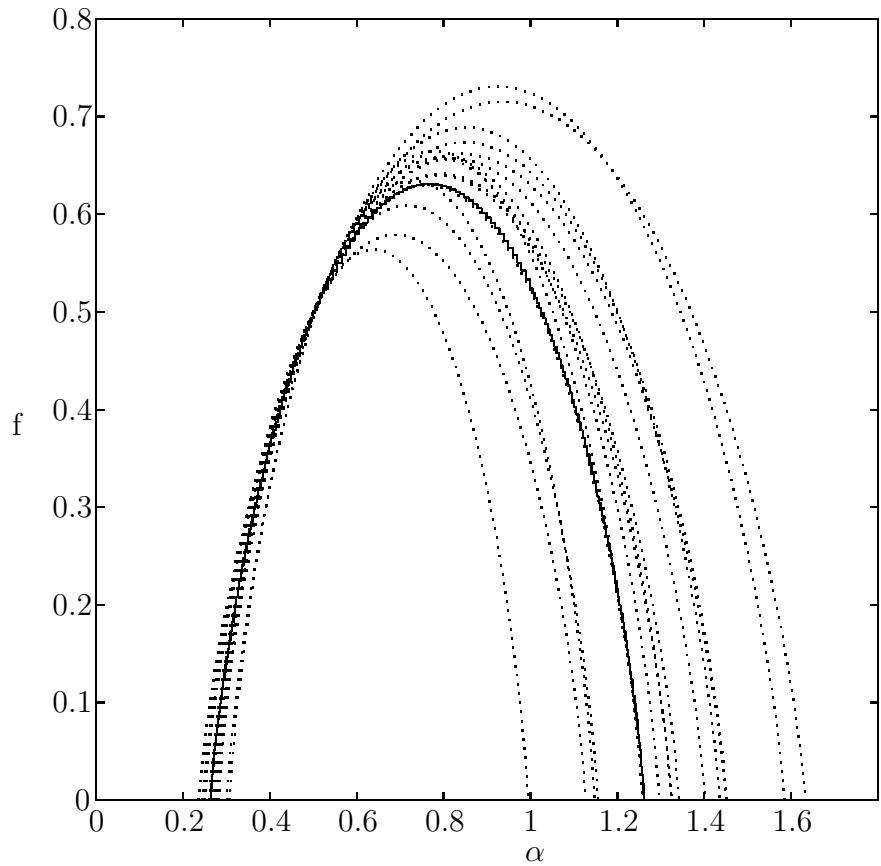

\setlength{\unitlength}{.07pt}

\caption{ensemble of multi-fractal spectra $f(\alpha)$, dotted, for each of
the predictions plotted in Figure~\protect\ref{n100means} made from samples
of $N=100$ data points.  For comparison the multi-fractal spectrum for the
actual fractal is plotted as the solid line.  Observe the good estimation
near the information dimension, but the large
errors for larger $\alpha$.}
\protect\label{n100fas}
\end{figure}

Ultimately, researchers want to examine multi-fractal properties of the
data.  In this method these will be determined from the parameters of the
best fit multi-fractal.  For example, for any determined parameters
$(\rho,\chi)$ for the binary multiplicative multi-fractal, the
multi-fractal spectrum of the data it matches would be predicted to be
given by the analytic expression~(\ref{emfs}).  For the 16 realisations of
the $N=100$ data-point multi-fractals, we determined various estimates,
plotted by $\circ$'s in Figure~\ref{n100means}, of the true parameter
values, $\rho=1/3$ and $\phi=1/4$.  For each of these estimates, we plot
the corresponding predicted $f(\alpha)$ curves in Figure~\ref{n100fas},
along with the true $f(\alpha)$ curve.

Observe that the predicted dimensions for low~$\alpha$, positive~$q$, are
quite good for all realisations, especially near the information dimension.
However, predicted dimensions for high~$\alpha$, negative~$q$, are poor;
this is also the case for the Hausdorff dimension which is predicted from
the maximum of the $f(\alpha)$ curve.

\section{Conclusion}

We have presented a method for predicting multi-fractal properties from
experimental data.  The technique is based on maximising the likelihood
that the given finite data set comes from a particular member of a
multi-parameter family of well known multi-fractals.  The evidence
indicates that properties predicted by the proposed method are unbiased by
the finite number of sample points in the experimental data.

A valuable feature of the method is that an indication of errors in the
best fit parameters is naturally obtained.  For example, the indications
are that predictions of the generalised dimensions $D_q$ for negative~$q$
are extremely unreliable.

Perhaps the method could be improved by increasing the resolution of
the correlation density in the $(\log r,\log c)$ plane.  However, this
would increase the number of ``bins'' of low but non-zero density which
may reduce the usefulness of the $\chi^2$ test.  Furthermore, it will
increase the correlations between the counts for each ``bin'' which may
will reduce the quality of the statistics, though perhaps not significantly.

Here we have addressed perhaps the simplest nontrivial example of the
process.  Most interesting data on spatial fractal distributions lie in two
or three dimensions where more complex families of multiplicative
multi-fractals need to be employed to model the characteristics of the data
distribution.  Here we only considered data in one dimension, but even then
we should propose a more adaptable approach by modelling the data through,
for example, a family of ternary or quarternary multiplicative
multi-fractals.  for example, such an enlarged family of comparison
multi-fractals will enable the method to match multi-fractals with
asymmetric $f(\alpha)$ curves.

Furthermore, here we knew the overall size of the fractal and that
there is no cut-off at small scales, or indeed that there is no
multi-scaling, in the data.  The technique will have to be developed
further to adapt to these features.


\addcontentsline{toc}{section}{References}
\begin{thebibliography}{10}

\bibitem{Borgani93}
S.~Borgani, G.~Murante, A.~Provenzale, and R.~Valdarnini.
\newblock Multifractal analysis of the galaxy distribution: reliability of
  results from finite data sets.
\newblock {\em Phys Rev E}, 47:3879--3888, 1993.

\bibitem{Emmerson95}
L.M. Emmerson and A.J. Roberts.
\newblock Fractal and multi-fractal patterns of seaweed settlement.
\newblock preprint, 1995.

\bibitem{Feder88}
J.~Feder.
\newblock {\em Fractals}.
\newblock Plenum Press, 1988.

\bibitem{Grassberger88b}
P.~Grassberger.
\newblock Finite sample corrections to entropy and dimension estimates.
\newblock {\em Phys Lett A}, 128:369--373, 1988.

\bibitem{Halsey86}
T.C. Halsey, M.H. Jensen, L.P. Kadanoff, I.~Procaccia, and B.I. Shraiman.
\newblock Fractal measures and their singularities: the characterization of
  strange sets.
\newblock {\em Phys Rev A}, 33:1141--1151, 1986.

\bibitem{Judd91}
K.~Judd and A.~I. Mees.
\newblock Estimating dimensions with confidence.
\newblock Technical report, University of Western Australia, 1991.

\bibitem{Mandelbrot79}
B.~B. Mandelbrot.
\newblock Fractals: Form chance and dimension.
\newblock {\em J. Fluid Mech.}, 92:206-- 208, 1979.

\bibitem{Martinez91}
V.J. Martinez.
\newblock Fractal aspects of galaxy clustering.
\newblock In Heck and Perdang, editors, {\em Applying fractals in astronomy},
  Lect Notes in Phys m3. Springer-Verlag, 1991.

\bibitem{Paladin87}
G.~Paladin and A.~Vulpiani.
\newblock Anomalous scaling laws in multifractal objects.
\newblock {\em Phys Rep}, 156:148--225, 1987.

\bibitem{Press92}
W.H. press, S.A. Teukolsky, W.T. Vetterling, and B.P. Flannery.
\newblock {\em Numerical recipes in FORTRAN. The art of scientific computing}.
\newblock CUP, 2nd edition, 1992.

\bibitem{Procaccia88a}
I.~Procaccia.
\newblock Universal properties of dynamically complex systems:the organisation
  of chaos.
\newblock {\em Nature}, 333:618--623, 1988.
\newblock 16th June.

\bibitem{Sreenivasan91}
K.R. Sreenivasan.
\newblock Fractals and multifractals in fluid turbulence.
\newblock {\em Annu. Rev. Fluid Mech.}, 23:539--600, 1991.

\end{thebibliography}
\end{document}